# Analysing user sentiment data for architectural interior spaces

Mi Kyoung Kim

School of Architecture, Hanyang University, Seoul 04763, Republic of Korea; nnyang2mk@hanyang.ac.kr (M.K.K.);
Tel.: +82-2-2220-1796

**Abstract:** This study aims to develop a data-driven system to enhance the analysis and improvement of user experiences in interior spaces, acknowledging the significant impact of design on individuals' health, productivity, and quality of life. Augmenting traditional methodologies such as surveys, interviews, and observations, the proposed approach integrates advanced technologies like LiDAR, UWB, and EEG to gather and analyze real-time user interactions. The objective is to monitor user locations and reactions in real time, evaluate emotional data via the ChatGPT API, and visualize this information on a spatial analysis grid using heatmaps. This comprehensive understanding of user perception, experience, and response within architectural spaces is anticipated to support architects and interior designers in creating user-centered spaces that enhance living conditions in smart buildings and urban planning. The study utilizes the 2022 Shanghai Jiaotong University Emotion EEG Dataset (SEED) for emotional response analysis, deliberately excluding physical environmental and socio-cultural factors from its scope.

## 1. Introduction

*1.1. Background*

Nowadays, the design of interior spaces is gaining more and more importance in shaping people's lives and experiences. This is due to the fact that individuals spend most of their time indoors, and the interior can have a significant impact on their health, productivity, and quality of life. To create spaces that meet the needs and preferences of users, it is necessary to have a comprehensive understanding of their behaviours, emotions, and perceptions in indoor environments.

User experience analyses for indoor spaces have been conducted using methodologies such as surveys, interviews, and observations. However, these methods may be limited in collecting and analysing direct information about users' feelings and reactions.

As a complementary method, a data-driven approach can be utilised. This is a method of gathering real-time information on users' interactions with their environment, utilising technologies such as LiDAR scanners, ultra-wideband (UWB) modules, and brainwave data collection systems.

The objective of this study is to develop an integrated data-driven system that can comprehensively analyse user perception, experience, and response to indoor architectural spaces. To achieve this, the study combines technologies such as LiDAR scanner, Unity3D game engine, ESP32 UWB, and EEG data acquisition system. The EEG equipment used is the Enobio EEG 32 device, which is certified as a Class IIa medical device with CE marking under Council Directive 93/42/EEC on medical devices (EC certificate ES19/86968) and licensed as a Class 2 medical device under the Canadian Medical Device Regulations, SOR 98/282 (licence number: 90344).  The aim is to monitor user location and reactions in real-time, analyse sentiment information using the ChatGPT API, and visualise it as a heatmap on an in-space analysis grid. This can aid in the effective design of indoor spaces and user experiences, and will assist architects and interior designers in making decisions to create user-centred spaces that contribute to improving quality of life in smart buildings and urban planning.

*1.2 Research Methods*

The aim of this research is to create an integrated data-driven system that can comprehensively analyse user perception, experience, and response to indoor architectural spaces. The research methodology involves collecting 3D point data of the indoor space and recreating it in virtual space using the Unity3D game engine.

Additionally, an ESP32 UWB-based user location recognition system was developed to analyse user movements in indoor spaces.

Real-time EEG data is collected from users in space using an Enobio EEG 32-channel device. The data is then analysed for emotional responses to the indoor environment using the LSL protocol.

To facilitate clear and easy interpretation of the emotional response of users in a space, heatmaps are used for visualization.

Comprehensive data analysis is performed by combining the ChatGPT API with Wolfram's Mathematica API. The combination of ChatGPT's text analytics and Mathematica's mathematical computing power reveals the effectiveness of user experience and design elements.

*1.3 Scope of the study*

This research focuses on user analytics and visualisation in indoor spaces. The scope of our research is limited to indoor spaces.

We use Electroencephalogram (EEG) data from the 2022 Shanghai Jiaotong University Emotion EEG Dataset (SEED) for user sentiment analysis.

We do not consider environmental factors. The user experience is not fully considered as the physical environmental factors, such as temperature, humidity, and lighting, are not taken into account.

Additionally, sociocultural factors, including the user's background, age group, and social context, are not considered.

**2. theoretical considerations**

*2.1 User experience in indoor spaces*

The research topic of user experience in indoor spaces has been of great interest to the fields of architecture, interior design, and environmental psychology. The aim is to understand how users perceive and interact with indoor spaces to create environments that are tailored to their needs and preferences, ultimately improving their health, productivity, and overall satisfaction. Various aspects of indoor space perception and user experience have been explored, including spatial configuration, colour, lighting, acoustics, and furniture arrangement. These factors can influence users' emotions, cognitive processes, and behaviour in the space. Studies have shown that well-designed spaces with appropriate lighting, colour schemes, and furniture arrangements can improve health by reducing stress and increasing satisfaction.

Recent technological advancements have enabled new approaches to researching indoor spatial perception and user experience. Virtual reality (VR) and augmented reality (AR) technologies can simulate and manipulate indoor environments, providing researchers with a controlled and immersive environment to explore user experience. It is important to note that these technologies should be used with objectivity and without bias. Data-driven methods have been used to analyse user experience objectively. Tools such as eye tracking, physiological measurements, and motion tracking have been employed.

Despite extensive research into indoor space perception and user experience, there is a lack of comprehensive data-driven systems that can capture and analyse users' positional awareness, emotions, and reactions in real-time. The integration of technologies, such as LiDAR scanners, the Unity3D game engine, ESP32 UWB, and EEG data acquisition systems, can provide valuable insights into effective design principles for indoor spaces.

*2.2 Perception of architectural space*

Light detection and ranging (LiDAR) scanners are a valuable tool for capturing accurate and detailed three-dimensional (3D) data of built spaces. By using laser pulses to measure distance, LiDAR scanners can generate high-resolution point clouds that represent the geometric surface characteristics of indoor and outdoor environments. LiDAR technology is being used for a variety of tasks, including building documentation, historic preservation, and remodelling projects.

LiDAR scanners are also used to analyse and understand user experience in indoor spaces. By creating accurate 3D models of the built environment, LiDAR data can be combined with the virtual reality (VR) or game engine platform Unity3D to simulate immersive interactions. This approach provides data to study user behaviour, emotions, and perceptions without requiring physical changes to the real space.

This study proposes an integrated data-driven system for capturing and analysing user experience in indoor architectural spaces, using a LiDAR scanner. The system collects detailed 3D point data of the indoor environment and implements it in the Unity3D game engine to provide an accurate and immersive spatial representation. This representation enables real-time analysis of user positional awareness, emotions, and reactions.

*2.3 Recognising the user's location*

Ultra-Wideband (UWB) technology offers precise, real-time location data for various applications. UWB is a wireless communication technology that uses a broad frequency range (usually above 500 MHz) and low power spectral density to provide high-accuracy ranging and positioning. This makes UWB an excellent option for indoor location systems, where GPS, Wi-Fi, or Bluetooth performance is reduced due to complex environments such as walls, furniture, and users.

In recent years, Ultra-Wideband (UWB) systems have been widely used in various indoor location awareness applications, such as navigation aids. These systems employ UWB anchors and UWB tags to track the user's location in real-time, and send this information to a server for processing.

This study uses an ESP32 UWB module to develop a user localisation system in an indoor architectural space. Accurately tracking the user's movements and interactions with the space can provide data on how the user experiences the indoor environment. By combining UWB technology with other components, such as the Unity3D game engine and an EEG data collection system, it is possible to conduct a comprehensive real-time analysis of the user's positional awareness, emotions, and reactions. This system can reflect the user's movements in the indoor environment in real-time in the virtual world, as illustrated in Figure 1.

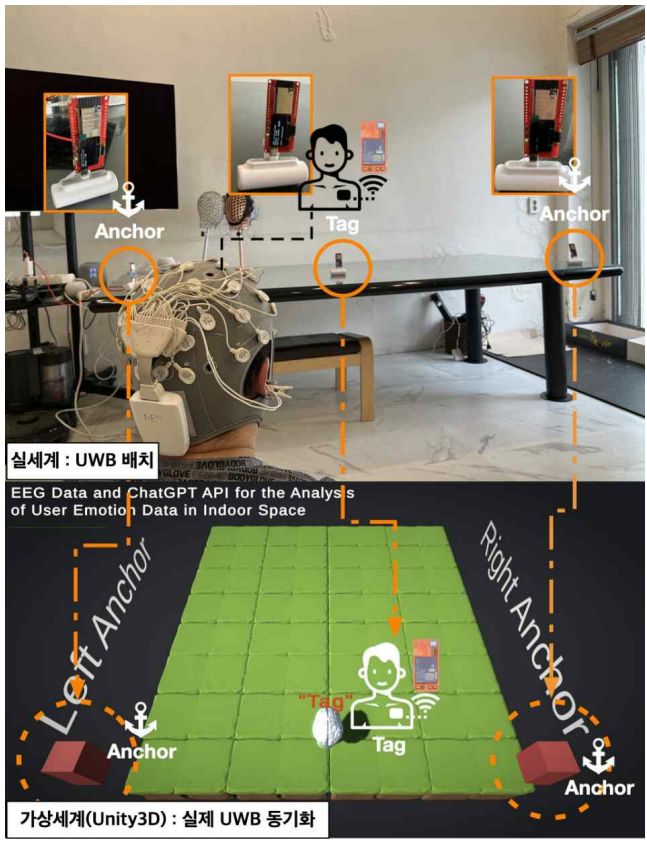

Figure 1. Convergence of real and virtual worlds: Implementation of a location awareness system using UWB

*2.4 EEG data and emotion analysis*

The analysis of electroencephalogram (EEG) data has become a significant method for studying human emotions, cognition, and behaviour. EEG measures the electrical activity of the brain through electrodes attached to the scalp, which can aid in understanding the neural processes involved in emotional and cognitive states without the need for experiments, using artificial intelligence.

This study uses the SEED dataset, developed by Shanghai Jiaotong University in China, which contains EEG data recorded while participants watched emotionally evocative video clips. The dataset has been employed in various fields, including emotion recognition, brain-computer interface (BCI) research, and brain signal processing. To construct an emotion model based on EEG, we utilized the SEED dataset, which has been collected and distributed by BMIS of Shanghai Jiaotong University since 2013 and is updated periodically. For this study, we used the 2022 version of the SEED dataset. Each participant in the SEED dataset watched 15 film clips three times while their EEG was recorded. The EEG was recorded using 62 channels and a sampling frequency of 200 Hz. Various methodologies have been used in studies utilizing this dataset to enhance the accuracy of emotion recognition. Wang et al. (2019) distinguished emotional states by using the PLV connectivity of EEG signals, achieving a classification accuracy of 84.35% on the SEED dataset. Li et al. (2021) proposed a Transferable Attention Neural Network (TANN) to achieve an accuracy of 84.41% on the SEED dataset. The studies demonstrate the significance of the SEED dataset in EEG-based emotion recognition research.

The study employed the change point detection technique to analyze EEG data for emotion recognition. This technique identifies points in time series data where the distribution or characteristics of the data change. An artificial intelligence model was developed to analyze EEG data in real-time and classify positive and negative emotions.

Enobio EEG 32-channel devices are used to monitor and evaluate users' emotional responses in real-time in indoor architectural spaces. This information is integrated with other components of the data-driven system, such as UWB-based location awareness and the Unity3D game engine, to enable effective analysis of the interior space and user experience.

Table 2. Procedure for conducting EEG sentiment analysis

| Steps | Contents |
| --- | --- |
| Prepare your data | Load the SEED dataset. |
| Preprocessing | Preprocess EEG data with data filtering, denoising, and data normalisation |
| Feature extraction | Feature extraction from EEG data using electrode-frequency distribution maps (EFDMs) based on short-time Fourier transforms |
| Building a model | Build a model for sentiment analysis, including setting up a convolutional neural network (CNN) with four residual blocks. |
| Training | Train a model on preprocessed data and extracted features |
| Evaluation | Evaluate the performance of your sentiment classification model |
| Predictions | Use trained models to analyse emotions in new EEG data |

### 3. Build a user sentiment data analysis system

*3.1 Collecting and processing indoor space data*

*3.1.1 Analysing real-world spatial data*

LiDAR scanners are utilised to collect high-resolution 3D point data of indoor architectural spaces.

The scanning process accurately captures indoor environments using laser light to measure distance and create precise, detailed 3D representations of objects and environments.

To perform the scanning process, place the LiDAR scanner in a designated location. The scanner emits laser pulses and measures the time it takes for the light to reach and return from the object, resulting in a dense point cloud of the indoor environment. The acquired point cloud data from the scan locations is registered and aligned to create a single, consistent 3D representation of the room. It is important to verify the accuracy of the registered point cloud data to ensure that all relevant areas of the indoor environment are adequately represented.

The point cloud data is collected, processed, and transformed into a format suitable for integration with the Unity3D game engine. This enables the creation of a virtual environment for real-time user experience analysis.

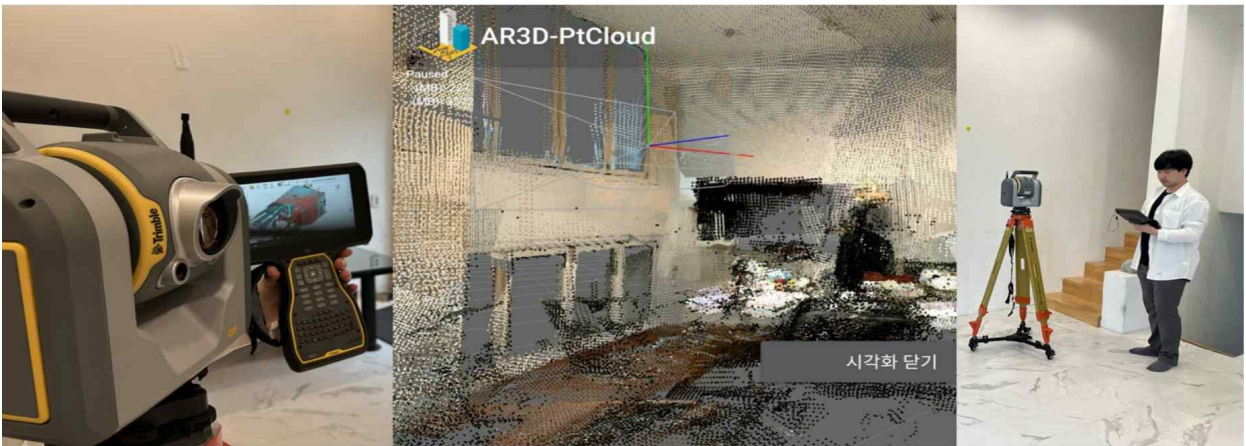
Figure 2. Indoor space scanning with Trimble SX12

*3.1.2 Analysing Virtual World Spatial Data*

After collecting and processing the LiDAR point cloud data of the indoor architectural space, it is implemented in the Unity3D game engine. Unity3D is an ideal platform for visualising and analysing user experiences in indoor spaces as it can create interactive 3D environments. The following steps describe the process of implementing LiDAR data in the Unity3D game engine:

Data conversion: Convert the processed LiDAR point cloud data or 3D mesh into a format compatible with Unity3D, such as OBJ, FBX, or Unity's native Asset format. This step involves using a third-party tool, Point Cloud Viewer and Tools, to assist in the conversion process.

Import the converted 3D model into your Unity3D project, ensuring that the model is correctly sized and oriented within the virtual environment.

Additionally, ensure that textures and materials are appropriately applied. To create a realistic representation of the interior space, we applied appropriate textures and materials to the 3D model. We used photos taken during the LiDAR scanning process and selected suitable textures from online libraries.

Lighting should be set up in the virtual environment to accurately represent the lighting conditions of a real-world interior space.

This involves setting up directional, point, or ambient lights, and adjusting light intensity, colour, and shadows. Additionally, an ESP32 UWB-based system should be integrated into the Unity3D environment to implement user location awareness in indoor spaces. The task involves developing an ESP32 UWB system as a script or plugin that can receive and process user location data, updating the user's location within the virtual environment in real-time.

To enable user interaction and navigation, implement user controls and navigation within the virtual environment.

Additionally, ensure that data visualisation is included. Visualisation tools, such as heatmaps and graphs, can be developed to display users' emotional responses and eye positions within a virtual environment.

By integrating LiDAR data into the Unity3D game engine, it is possible to create a virtual representation of an indoor architectural space to analyse user experience and emotions in real-time. This interactive experience forms the basis of a data-driven system that integrates user location tracking, brainwave data analysis, and data visualisation to gain a better understanding of the factors that contribute to a positive user experience in indoor spaces.

*3.2 User Location Recognition System*

The proposed data-driven system for indoor space recognition and user experience analysis relies heavily on accurately recognizing the user's location in the architectural space and collecting biometric data corresponding to those coordinates. To achieve this, we have developed an ESP32 ultra-wideband (UWB) based location recognition system.

Please refer to Table 1 for the process of developing and integrating this system.

Table 1. Components of the ESP32 UWB

| Anchor | TAG | UWB calculation demo |
|---|---|---|
| 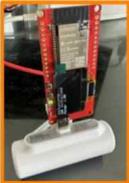 | 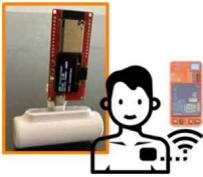 | 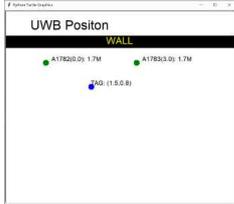 |
| Track user location after pinning in space | Real-time location tracking after user possession | Green dot: Anchor<br>Blue dot: Tag user location |

Hardware Selection and Configuration: To ensure optimal performance of your device, it is important to carefully select and configure the appropriate ESP32 UWB module. We recommend using the Makerfabs ESP32 UWB Module. When configuring the UWB module, consider your project requirements and compatibility with the room environment and other system components.

Anchor Node Placement: To optimize positioning accuracy and coverage, it is recommended to place multiple UWB anchor nodes within the room, taking into account the size and layout of the room.

UWB Positioning Algorithm: This task involves developing a positioning algorithm that estimates the user's location in an indoor space based on time-of-arrival (ToA) or time-difference-of-arrival (TDoA) measurements obtained from the UWB module. The positioning algorithm options include trilateration, multilateration, or particle filtering.

Integration with Unity3D: The goal is to create a script or plugin within the Unity3D game engine that can receive and process user location data from the ESP32 UWB system. The script updates the user's location in the virtual environment in real-time, based on the estimated location data.

We have constructed and integrated a cube-based analytics grid for collecting biometric data within the game engine environment. This grid is created in a cube structure and serves as the basis for mapping the location and biometric data of virtual users.

It is important to calibrate and test the system. The objective of this task is to calibrate the ESP32 UWB-based positioning system to ensure accurate and consistent position estimation within indoor spaces.

Extensive testing should be performed to verify the reliability and accuracy of the system under various conditions and scenarios.

The ESP32 UWB system implements a real-time data transfer protocol to continuously stream location data to the Unity3D game engine using wireless communication technologies such as Wi-Fi or Bluetooth.

The proposed data-driven system integrates an ESP32 UWB-based location awareness system to accurately track and analyse user location within indoor architectural spaces. This location data can be combined with EEG data analysis and data visualisation techniques to provide insights into the relationship between user experience and interior space design.

*3.3 Collecting and analysing real-time EEG data*

Collecting and analysing EEG data for user experience analysis is crucial. This data pertains to a user's emotional and cognitive state and can aid architects and designers in comprehending how the design of an interior space impacts the user experience. The subsequent step involves collecting and analysing EEG data in real-time.

To collect real-time EEG data from users, we use the Enobio EEG 32-channel system.

We adhere to the Lab Streaming Layer (LSL) protocol for EEG data collection. LSL is a protocol that facilitates the synchronization of data streams from multiple devices. It plays a crucial role in the proposed data-driven system. LSL enables the ESP32 UWB system to collect and analyze user location data and EEG data simultaneously (refer to Figure 1). This synchronization is essential for conducting a comprehensive analysis of user experience in indoor spaces. It allows for the examination of the relationship between user emotions and location.

Data preprocessing is a crucial step in ensuring the accuracy and reliability of sentiment analysis on collected EEG data. Data preprocessing is a crucial step in ensuring the accuracy and reliability of sentiment analysis on collected EEG data. This involves filtering, removing artefacts, and segmenting the data.

Emotion analysis is then performed using machine learning algorithms and emotion recognition models to identify the user's emotional state or cognitive response in real-time.

The proposed data-driven system collects and analyses EEG data in real-time to provide comprehensive information on the emotional and cognitive state of users in indoor architectural spaces.

Table 2. Classification of housing ownership

| lent | self-owned | total |
|---|---|---|

```python
import openai
import pandas as pd
import numpy as np
from sklearn.model_selection import train_test_split
from sklearn.preprocessing import MinMaxScaler
from transformers import GPT4Config, GPT4ForSequenceClassification, GPT4Tokenizer, Trainer, TrainingArguments

# 1. set API key
openai.api_key = "your-api-key"

# 2. prepare EEG data and emotion labels
# Read a CSV file containing EEG data and emotion labels
data = pd.read_csv('brainwave_data.csv')

# 3. Data preprocessing
# normalise data
scaler = MinMaxScaler()
data_normalised = scaler.fit_transform(data.iloc[:, :-1])

# split dataset
X_train, X_test, y_train, y_test = train_test_split(data_normalised, data.iloc[:, -1], test_size=0.2, random_state=42)

# 4. Fine-tuning the OpenAI GPT-4 model
config = GPT4Config.from_pretrained("gpt4-base", num_labels=len(set(y_train)))
tokenizer = GPT4Tokenizer.from_pretrained("gpt4-base")
model = GPT4ForSequenceClassification.from_pretrained("gpt4-base", config=config)

# Data tokenisation
train_encodings = tokenizer(X_train.tolist(), truncation=True, padding=True)
test_encodings = tokenizer(X_test.tolist(), truncation=True, padding=True)

# Create a Python dataset
train_dataset = openai.Dataset(train_encodings, y_train)
test_dataset = openai.Dataset(test_encodings, y_test)

# Set training factors
```

```
training_args = TrainingArguments(
    output_dir="./results",
    num_train_epochs=5,
    per_device_train_batch_size=8,
    per_device_eval_batch_size=8,
    evaluation_strategy="epoch",
    logging_dir="./logs",
)

# Training and assessment
trainer = Trainer(
    model=model,
    args=training_args,
    train_dataset=train_dataset,
    eval_dataset=test_dataset,
)
trainer.train()

# 5. Evaluate the results
predictions = trainer.predict(test_dataset)
predicted_labels = np.argmax(predictions, axis=1)

accuracy = np.sum(predicted_labels == y_test) / len(y_test)
print(f"Accuracy: {accuracy:.2f}")
```

*3.4 Visualising user sentiment data*

Visualise the user's emotional state and reactions in the Unity3D virtual environment, using a heatmap to display the user's emotions and location in relation to the room space. Colours represent different emotional states or intensities. The next step is to create the heatmap visualisation.

Effective visualisation of emotional responses: Heatmap visualisations represent users' emotional responses in an interior space. The text describes how displaying areas where positive or negative emotions are experienced can help identify specific design elements and spatial arrangements that influence user experience and emotion.

It also mentions the integration of Unity3D game engine with the ESP32 UWB-based location awareness system for heatmap visualisation. This integration enables the real-time visualisation of users' emotional responses and location in a virtual environment, providing a comprehensive understanding of the dynamic relationship between user experience, location, and architectural design.

The heatmap visualisation methodology offers customisable visualisations, allowing researchers to adjust parameters to suit the specific goals and needs of the research project. This enabled the exploration of various design scenarios and their impact on user emotional responses and eye position.

The colours in a heatmap visualisation, which represents different emotional states or intensities, allow architects and designers to identify areas within a space that generate positive or negative emotional responses. This provides a clear and intuitive suggestion for future design changes and improvements.

*3.5 Comprehensive data analysis using the ChatGPT API*

The ChatGPT API can be utilized to generate a comprehensive sentiment analysis based on the collected EEG data and user location information. To leverage the biometric information, the ChatGPT API can be used for decision making, while Wolfram can be used for mathematical computation.

The ChatGPT API is capable of processing and interpreting text-based data, enabling the identification of patterns and themes related to indoor space design and user experience. This text analysis can provide data on factors that influence users' emotions, perceptions, and preferences in indoor spaces.

The ChatGPT API successfully enhanced the data analysis process by providing natural language understanding and generation capabilities to help interpret and contextualise the data collected. This enabled a more comprehensive understanding of user experience and emotions in relation to interior space design and arrangement.

Pattern recognition and correlation: Researchers leveraged the ChatGPT API to identify patterns and correlations between user emotional responses, gaze positions, and specific design elements in interior spaces. These insights suggest effective design principles and strategies for optimizing user experience and health in built spaces.

The analytics approach is customizable. The ChatGPT API provides a high level of customisation and adaptability, enabling researchers to tailor their analytics approach to meet the specific needs and goals of their research project. This flexibility enables them to explore different design scenarios and evaluate the impact on user emotional response and experience.

Integration with other data sources is also possible. The integration of ChatGPT API-based insights with collected EEG data, heatmap visualisations, and ESP32 UWB-based location awareness data was successful. This integration provided a comprehensive understanding of the user experience in indoor spaces and supported the development of data-driven design principles for improving the user experience.

The scalability and applicability of these principles were also considered. The ChatGPT API-based insights demonstrated scalability and applicability. They enabled researchers to analyse large datasets and gain insights that can be applied to various architectural spaces and design scenarios. This diversity supports the generalisability of the findings and their relevance to different interior space and design contexts.

In conclusion, the ChatGPT API-based insights demonstrate the methodology's effectiveness in providing comprehensive data analysis and insights into user experience and emotions in indoor spaces. The integration of ChatGPT API with other data sources allows for customisation, scalability, and applicability. This integration can help researchers better understand the relationship between indoor space design and user experience, leading to the development of effective design principles that improve user health in built spaces.

Quantitative data insights: The ChatGPT API can also be used to analyse quantitative data such as EEG data, user location information, and eye gaze patterns to generate natural language descriptions and summaries of the data. This capability allows researchers to easily understand complex data patterns and relationships, which in turn provides valuable insights into what shapes user experience in indoor spaces.

Analyse across data types: By integrating the ChatGPT API into your data analysis process, you can examine qualitative and quantitative data simultaneously. This cross-analysis can reveal interactions between different factors, such as relationships between user sentiment, eye gaze patterns, and location data, ultimately providing a more comprehensive understanding of user experience in indoor spaces.

Data-driven recommendations: The ChatGPT API can generate data-driven recommendations and suggestions based on analytics results. These recommendations help architects and designers make informed decisions about the design and layout of interior spaces, contributing to the creation of more effective and user-centric built spaces that enhance user experience and improve overall quality of life.

Scalability and efficiency: By leveraging the ChatGPT API for data analysis, researchers can efficiently analyse large amounts of data and gain valuable insights in a timely manner. This scalability and efficiency is especially important for conducting comprehensive data-driven research on large indoor spaces or complex architectural designs.

In conclusion, the integration of ChatGPT API for indoor space perception and user experience analysis in the proposed data-driven system provides comprehensive data analysis and valuable insights into users' emotions, perceptions, and experiences in indoor spaces. By leveraging the capabilities of the ChatGPT API, researchers can better understand the factors that shape user experience in indoor spaces, which in turn can guide the development of more effective and user-centred architectural spaces.

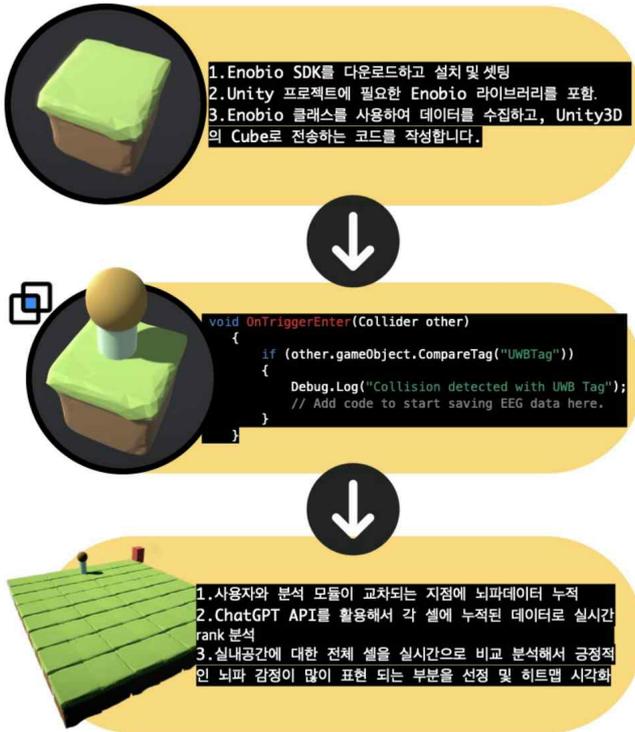

## 4. conclusion

*4.1 Summary of findings*

This research aimed to develop an integrated data-driven system for comprehensive analysis of user location perception, experience, and reactions in indoor architectural spaces. By utilising the LSL protocol with a LiDAR scanner, Unity3D game engine, ESP32 UWB and EEG data, the study succeeded in tracking and analysing user location and emotions in real-time.

Key findings from the study include

Successfully implement 3D point data of indoor architectural spaces in the Unity3D game engine to create accurate and detailed virtual representations of indoor environments.

ESP32 UWB-based systems can be effectively developed to provide accurate user localisation within indoor spaces to track user movement and interaction with the environment.

Collect and analyse user brainwave data in real-time to provide critical insights into user emotional states and experiences within indoor spaces.

Use heatmaps to visualise users' emotional responses to understand how different aspects of interior space design affect the user experience.

Comprehensive data analysis using the ChatGPT API provides critical insights and recommendations for user-centred interior space design.

These findings contribute to a deeper understanding of effective design principles for interior spaces and user experience, helping architects and interior designers create user-centred spaces that improve quality of life in future smart buildings and urban planning.

*4.2 Implications for architects and interior designers*

The findings have several implications for architects and interior designers, providing important insights for creating user-centred interior spaces:

Educated design decisions: By understanding how different aspects of an interior space (layout, lighting, and materials) affect user emotions and experiences, architects and interior designers can make more educated design decisions that meet users' needs and preferences.

User-centred approach: The importance of a user-centred approach in architecture and interior design is highlighted by the real-time analysis of user emotions and experiences within interior spaces. Designers should prioritise the needs and expectations of users in the design process to create spaces that enhance their well-being and satisfaction.

Optimise space utilisation: With an understanding of how users interact and move through interior spaces, architects and interior designers can optimise space utilisation and functionality. This can lead to better space utilisation, resulting in cost savings and an improved user experience.

Adaptability and flexibility: Insights from this research can help architects and interior designers create spaces that are adaptable and flexible to the changing needs of users over time. This promotes long-term user satisfaction and well-being.

Data-driven design: The data-driven approach used in this study highlights the importance of incorporating evidence-based design principles in architecture and interior design practice. By utilising real-time data and analytics, architects and interior designers can make better, more evidence-based decisions, leading to more effective design outcomes.

Overall, the findings provide architects and interior designers with important insights and guidance for creating user-centred interior spaces. By embracing a data-driven approach and prioritising user emotions and experiences, designers can contribute to improving quality of life in future smart buildings and urban planning.

*4.3 Contribution to improving quality of life*

This research provides several contributions to improving the quality of life in future smart buildings and urban planning through the development of user-centred indoor spaces:

Enhanced user well-being: By focusing on user emotions and experiences, architects and interior designers can create interior spaces that promote user well-being, comfort, and satisfaction. This, in turn, can contribute to improved mental and physical health, increased productivity, and user happiness.

Personalisation: Real-time analysis of user experience and emotions allows architects and interior designers to tailor interior spaces to the needs and preferences of individual users. Personalisation of spaces can increase user satisfaction and strengthen a sense of belonging, improving overall quality of life.

Inclusive design: By understanding the different needs and experiences of users, architects and interior designers can create inclusive spaces that accommodate a wide range of users, including people with disabilities, the elderly, and children. Inclusive design can contribute to a fairer and more inclusive society by ensuring that everyone can access, use, and enjoy interior spaces.

Sustainable and energy-efficient spaces: The data-driven approach used in this study can also inform the design of sustainable and energy-efficient interior spaces. By optimising space utilisation, materials and lighting, architects and interior designers can reduce energy consumption and the overall environmental impact of buildings, contributing to a sustainable future.

Community building and social interaction: By designing interior spaces that encourage social interaction and create a sense of community, architects and interior designers can contribute to the development of strong social networks and relationships among users. This helps to promote a sense of belonging and support within a community, improving overall quality of life.

In conclusion, this research contributes to improving quality of life in future smart buildings and urban planning by providing important insights and guidelines for creating user-centred indoor spaces. By focusing on user emotions and experiences, architects and interior designers can improve the quality of life for all users by creating environments that promote well-being, inclusivity, sustainability, and community building.

**Author Contributions:**

**Funding:** This work was supported by a National Research Foundation of Korea (NRF) grant funded by the Korean government. (2021R1C1C2012502)

# References

1. Alalouch, C., Aspinall, P., & Smith, H. (2016). Using virtual reality to investigate user experience in indoor spaces. In Proceedings of the 3rd International Conference on VR Technologies in Cultural Heritage (pp. 1-8).
2. Getting Started with ESP32 DW1000 UWB (Ultra Wideband) Module. How To Electronics. https://how2electronics.com/getting-started-with-esp32-dw1000-uwb-ultra-wideband-module/
3. M. Aljehani et al. "A 3D point cloud-based indoor mapping system using a 2D laser scanner," IEEE Access, vol. 8, pp. 15367-15379, 2020.
4. Y. Li et al. "An ESP32 UWB-based indoor localisation system," IEEE Internet of Things.
5. Gannouni, S., Aledaily, A., Belwafi, K., et al. Emotion detection using electroencephalography signals and a zero-time windowing-based epoch estimation and relevant electrode identification. Sci Rep, 11, 7071. https://doi.org/10.1038/s41598-021-86345-5
6. Kim, B.J., Kim, H.J., & Ha, Y.C. (2014). A study on the net pressure coefficient distributions of ginseng facilities for wind resistant design - Focus on partially opened and closed wall type, Journal of the Wind Engineering Institute of Korea, 18(3), 163~172.
7. Korea Research Institute For Human Settlements (1988). A Case Study on Seoul Housing Market, 34.
8. Lee, C. S. (2007). Differences in housing tenureship by the characteristics of household heads, Journal of the Architectural Institute of Korea, Planning and Design Section, 23(2), 119-127.
9. Lerman, S. (1975). A Disaggregate Behavioural Model of Urban Mobility Decisions, Ph.D. Dissertation, M.I.T.
10. Luce, R. (1959). Individual Choice Behaviour, New York, John Wiley & Sons, 138-140.